\documentclass[%
 aip,
 amsmath,amssymb,
preprint,%
]{revtex4-1}

\usepackage{graphicx}% Include figure files
\usepackage{dcolumn}% Align table columns on decimal point
\usepackage{bm}% bold math
\usepackage[utf8]{inputenc}
\usepackage[T1]{fontenc}
\usepackage{mathptmx}
\usepackage{etoolbox}

\usepackage{xcolor}
\usepackage{hyperref}

\makeatletter
\def\@email#1#2{%
 \endgroup
 \patchcmd{\titleblock@produce}
  {\frontmatter@RRAPformat}
  {\frontmatter@RRAPformat{\produce@RRAP{*#1\href{mailto:#2}{#2}}}\frontmatter@RRAPformat}
  {}{}
}%
\makeatother
\begin{document}
% \preprint{JCP22-AR-02991}

% \title[Deep Coarse-grained Potentials via Relative Entropy Minimization]{Deep Coarse-grained Potentials via Relative Entropy Minimization}
\title[]{Deep Coarse-grained Potentials via Relative Entropy Minimization}
% Force line breaks with \\
\author{Stephan Thaler}

\author{Maximilian Stupp}

\author{Julija Zavadlav}
\email{julija.zavadlav@tum.de}
\altaffiliation[Also at ]{Munich Data Science Institute $\&$ Munich Institute for Integrated Materials, Energy and Process Engineering, Technical University of Munich, Germany}%Lines break automatically or can be forced with \\
\affiliation{%
Professorship of Multiscale Modeling of Fluid Materials,
Department of Engineering Physics and Computation,
TUM School of Engineering and Design, 
Technical University of Munich, Germany
}%

\date{\today}% It is always \today, today,

\begin{abstract}
    Neural network (NN) potentials are a natural choice for coarse-grained (CG) models. Their many-body capacity allows highly accurate approximations of the potential of mean force, promising CG simulations at unprecedented accuracy.
    CG NN potentials trained bottom-up via force matching (FM), however, suffer from finite data effects: They rely on prior potentials for physically sound predictions outside the training data domain and the corresponding free energy surface is sensitive to errors in transition regions.
    The standard alternative to FM for classical potentials is relative entropy (RE) minimization, which has not yet been applied to NN potentials.
    In this work, we demonstrate for benchmark problems of liquid water and alanine dipeptide that RE training is more data efficient due to accessing the CG distribution during training, resulting in improved free energy surfaces and reduced sensitivity to prior potentials. In addition, RE learns to correct time integration errors, allowing larger time steps in CG molecular dynamics simulation while maintaining accuracy. Thus, our findings support the use of training objectives beyond FM as a promising direction for improving CG NN potential accuracy and reliability.
\end{abstract}

\maketitle

\section{Introduction}
Molecular dynamics (MD) simulations are a popular tool for studying bio-physical processes at the nanoscale. For atomistic (AT) simulations, time and length scales of many processes of interest are still out of reach on currently available computational hardware. Coarse-graining \cite{Mccoy1998, Reith2003, Marrink2007, Noid2008, Shell2008, Noid2013, Ingolfsson2014, Singh2019} (CG) - grouping AT particles into effective interaction beads - is a common approach to model these systems as larger spatiotemporal scales can be reached due to a reduced number of interactions and an increased time step size \cite{Ingolfsson2014}.

The fidelity of CG simulations strongly depends on the employed CG potential energy function that defines particle interactions.
In the classical CG literature, potentials follow simple functional forms \cite{Marrink2007, Ingolfsson2014}. Recent years have seen an increased use of neural network (NN) potentials \cite{Behler2007, Behler2011, Schutt2017, Gilmer2017, Klicpera2020, Klicpera2020b, Qiao2020, Jain2021, Ko2021, Batzner2022, Batatia2022} for CG models - both for bottom-up learning to match properties of AT models \cite{Zhang2018a, Wang2019, Loeffler2020, Husic2020, Chen2021, Ding2022, Kohler2022} and for top-down learning to match experimental data \cite{Thaler_2021}.
In the following, we focus on the bottom-up learning case with the aim to obtain a CG model that is consistent with an existing AT model.
Consistency is achieved if the distribution of CG states sampled from the CG model equals the distribution generated by the AT model when mapping the AT states to CG coordinates \cite{Noid2008}.
In this case, the CG potential equals the many-body potential of mean force (PMF) \cite{Noid2008}. Consequently, NN potentials are a natural choice for CG potential energy functions: Their many-body nature \cite{Batatia2022b} allows for a more accurate approximation of the PMF than classical CG models, promising CG simulations at unprecedented accuracy.

So far, most bottom-up CG NN potentials have been trained via force matching (FM) \cite{Zhang2018a, Wang2019, Loeffler2020, Husic2020, Chen2021}.
FM minimizes the difference between CG force predictions and corresponding target AT forces for a given data set \cite{Noid2008, Wang2019}, typically generated by an AT MD simulation. FM training, while computationally inexpensive and straightforward to implement, suffers from two problems caused by the low availability of high energy states \cite{Herr2018}.
First, reproducing the ratio of different meta-stable states proves difficult for CG NN potentials trained via FM \cite{Chen2021, Kohler2022}: The CG potential is thought to be susceptible to errors in rarely sampled transition regions that affect the global accuracy of the free energy surface (FES) \cite{Kohler2022}.
Second, NN potentials are physics-free universal function approximators. Thus, they heavily rely on prior potentials that enforce qualitatively correct force predictions outside the training data distribution to avoid unphysical states, e.g. particle overlaps \cite{Wang2019, Husic2020}. Both of these problems can cause erroneous results in subsequent CG MD simulations, but critically, their extent is not reflected in the FM validation error during training \cite{Kohler2022}.

Given these drawbacks of FM in practice, recent efforts focused on training schemes beyond FM, including noise-contrastive estimation \cite{Gutmann2010, Ding2022} and flow-matching \cite{Kohler2022}.
Another alternative to FM is relative entropy (RE) minimization \cite{Shell2008}, which has been frequently used to optimize classical CG models \cite{Carmichael2012, Bottaro2013, Mashayak2015, Sanyal2018}, but has not yet been applied to CG NN potentials.
The main conceptual difference between RE and FM lies in the availability of molecular states during training: While FM trains exclusively on states provided by the AT model, RE additionally samples states from the CG model \cite{Shell2008, Chaimovich2011}. Sampling the CG model at each update step is computationally expensive, but it gives direct access to the CG distribution during training. Thus, deviations from the AT distribution can be accessed and subsequently corrected via gradient descent optimization - subject to the functional form of the CG model and the statistical sampling error.
Thus, in the context of CG NN potentials, RE counters suboptimal global FESs and sensitivity to prior potentials.

In this work, we demonstrate the effectiveness of optimizing CG NN potentials via RE minimization.
To this end, we train the CG DimeNet++ \cite{Klicpera2020, Klicpera2020b} graph NN potential for the benchmark problems of liquid water and alanine dipeptide. For liquid water, both FM and RE yield highly accurate CG potentials, but RE allows larger time steps in subsequent CG MD simulations without compromising accuracy. For alanine dipeptide, the RE method results in a more accurate FES and is more robust to the choice of prior potential compared to FM. Finally, we showcase that pre-training via FM allows to reduce the computational cost of RE training. Hence, the exploitation of training targets beyond FM is a promising path towards next generation CG NN potentials.

\section{Methods}
We reiterate fundamentals of FM \cite{Izvekov2005, Noid2008, Noid2008b, Mullinax2009} and RE minimization \cite{Shell2008, Chaimovich2009, Chaimovich2010, Chaimovich2011, Espanol2011, Shell2016} theory, based on which we discuss specific properties of both methods in the context of CG NN potentials.
The starting point for CG modeling is the selection of a mapping function $\mathbf{M}$
\begin{equation}
    \mathbf{R} = \mathbf{M}(\mathbf{r}) \ ,
\end{equation}
which maps AT coordinates $\mathbf{r} \in \mathbb{R}^{3n}$ onto a lower-dimensional set of CG coordinates ${\mathbf{R} \in \mathbb{R}^{3N}}$ with $N < n$. In the following, we assume canonical (NVT) ensembles in equilibrium and $\mathbf{M}$ to be a linear function, even though generalizations to non-equilibrium systems \cite{Harmandaris2016} and non-linear mappings \cite{Kalligiannaki2015} exist.

The CG model is consistent with the underlying AT model if the configurational equilibrium distribution of the CG model $p^\mathrm{CG}_{\bm \theta}(\mathbf{R})$ equals $p^\mathrm{AT}(\mathbf{R})$, the configurational equilibrium distribution of the AT model $p^\mathrm{AT}(\mathbf{r})$, when mapped to CG coordinates \cite{Noid2008}
\begin{equation}
    p^\mathrm{AT}(\mathbf{R}) = \left \langle \delta[\mathbf{R} - \mathbf{M}(\mathbf{r})] \right \rangle_\mathrm{AT} \ ,
\end{equation}
where $\langle ... \rangle_\mathrm{AT}$ indicates an AT ensemble average.
$p^\mathrm{CG}_{\bm \theta}(\mathbf{R})$ depends on model parameters $\bm \theta$ via the CG potential $U^\mathrm{CG}_{\bm \theta}(\mathbf{R})$.
The CG model is consistent with the AT model if the CG potential equals the many-body potential of mean force (PMF) \cite{Noid2008, Shell2008}
\begin{equation}
    U^\mathrm{PMF}(\mathbf{R}) = - \frac{1}{\beta} \ln p^\mathrm{AT}(\mathbf{R}) + C \ ,
\end{equation}
where $\beta = 1 / (k_\mathrm{B} T)$ with Boltzmann constant $k_\mathrm{B}$ and temperature $T$. $C$ is an arbitrary constant that we omit in the following.
To approximate the PMF, the most popular methods are the FM \cite{Izvekov2005, Noid2008, Noid2008b} and the RE minimization \cite{Shell2008} method.

\subsection{Force Matching}
FM - also known as multiscale coarse-graining \cite{Izvekov2005, Noid2008, Noid2008b} - aims to match the CG forces $ - \nabla_\mathbf{R} U^\mathrm{CG}_{\bm \theta}(\mathbf{R})$ to the instantaneous forces acting on CG particles $\mathbf{F}^{AT}$ computed from the AT system.
Thus, FM minimizes the mean squared error (MSE) loss function
\begin{equation}
\label{eq:FM_loss}
    \chi^2(U^\mathrm{CG}_{\bm \theta}) = \left \langle ||\mathbf{F}^\mathrm{AT}  + \nabla_\mathbf{R} U^\mathrm{CG}_{\bm \theta}(\mathbf{M}(\mathbf{r}))||^2  \right \rangle_\mathrm{AT} \ ,
\end{equation}
where $||...||$ is the Frobenius norm.
In practice, the AT ensemble average is approximated by the mean over a reference data set of AT configurations, typically generated by an AT MD simulation \cite{Noid2008b}.
Minimizing the loss in eq. \ref{eq:FM_loss} represents a standard supervised learning problem, which is solved by computing the gradient $\nabla_{\bm \theta} \chi^2(U^\mathrm{CG}_{\bm \theta})$ via automatic differentiation for a mini-batch of AT configurations and updating $\bm \theta$ via a stochastic optimizer \cite{Wang2019}.

To connect $U^\mathrm{CG}_{\bm \theta}$ to the PMF, eq. \ref{eq:FM_loss} can be reformulated \cite{Noid2008} as
\begin{equation}
\label{eq:FM_PMF_form}
    \chi^2(U^\mathrm{CG}_{\bm \theta}) = \langle ||\nabla_\mathbf{R} U^\mathrm{CG}_{\bm \theta}(\mathbf{M}(\mathbf{r})) - \nabla_\mathbf{R} U^\mathrm{PMF}(\mathbf{M}(\mathbf{r}))||^2\rangle_\mathrm{AT} + \underbrace{\langle ||\mathbf{F}^\mathrm{AT} + \nabla_\mathbf{R} U^\mathrm{PMF}(\mathbf{M}(\mathbf{r}))||^2  \rangle_\mathrm{AT}}_{\equiv \chi^2(U^\mathrm{PMF})} \ .
\end{equation}
Note that $\chi^2(U^\mathrm{PMF})$ depends on the CG mapping $M$, but cannot be optimized via $\bm \theta$.
Thus, FM minimizes the first term, resulting in the force predictions of the CG potential to approximate the forces of the PMF.
For infinite data and model capacity, $U^\mathrm{CG}_{\bm \theta}$ therefore converges to $U^\mathrm{PMF}$ (up  to an additive constant). From a ML perspective, the second term in eq. \ref{eq:FM_PMF_form} corresponds to the noise term in a regression problem \cite{Wang2019}. Physically, the noise term results from the fact that multiple AT states with different $\mathbf{F}^{AT}$ map to the same CG configuration. Hence, the noise term is irreducible and constitutes the lower bound of the loss.

\subsection{Relative Entropy Minimization}
The relative entropy - known as Kullback-Leibler divergence \cite{Kullback1951} in information theory - is commonly used to quantify the distance between two distributions.
In the context of CG modeling, these two distributions are $p^\mathrm{AT}(\mathbf{R})$ and $p^\mathrm{CG}_{\bm \theta}(\mathbf{R})$, defining the relative entropy $S_\mathrm{rel}$ as \cite{Espanol2011, Rudzinski2011, Shell2016}
\begin{equation}
    \label{eq:CG_S_rel}
    S_\mathrm{rel}(U^\mathrm{CG}_{\bm \theta}) = \int p^\mathrm{AT}(\mathbf{R}) \ln \left( \frac{p^\mathrm{AT}(\mathbf{R})}{p^\mathrm{CG}_{\bm \theta}(\mathbf{R})} \right) \mathrm{d}\mathbf{R} \ .
\end{equation}
Due to Gibbs' inequality, $S_\mathrm{rel}(U^\mathrm{CG}_{\bm \theta}) \geq 0$. Consequently, $S_\mathrm{rel}(U^\mathrm{PMF}) = 0$ is the global minimum, reached if $p^\mathrm{AT}(\mathbf{R}) = p^\mathrm{CG}_{\bm \theta}(\mathbf{R})$ and $U^\mathrm{CG}_{\bm \theta} = U^\mathrm{PMF}$ \cite{Rudzinski2011}. Thus, minimization of $S_\mathrm{rel}(U^\mathrm{CG}_{\bm \theta})$ provides a means to approximate $U^\mathrm{PMF}$.

Inserting the configurational probabilities of the canonical ensemble into eq. \ref{eq:CG_S_rel} yields \cite{Shell2008, Chaimovich2011, Shell2016}
\begin{equation}
\label{eq:RE_definition_canonical}
    S_\mathrm{rel}(U^\mathrm{CG}_{\bm \theta}) = \beta \langle U^\mathrm{CG}_{\bm \theta}(\mathbf{M}(\mathbf{r})) - U^\mathrm{AT}(\mathbf{r}) \rangle_\mathrm{AT} - \beta (A_{\bm \theta}^\mathrm{CG} - A^\mathrm{AT}) + S_\mathrm{map} \ ,
\end{equation}
where $A$ is the Helmholtz free energy and $S_\mathrm{map}$ depends on the mapping function $M$, but is independent of $\bm \theta$ \cite{Chaimovich2011, Shell2016}.
For classical CG potentials, $S_\mathrm{rel}(U^\mathrm{CG}_{\bm \theta})$ is typically minimized via the Newton-Raphson scheme \cite{Shell2008, Chaimovich2009, Chaimovich2011, Bottaro2013}.
In this work, we follow standard deep learning practice and optimize the NN potential via a first order optimizer. This approach avoids the high memory cost of computing the Hessian of the NN parameter set.
While computing $S_\mathrm{rel}(U^\mathrm{CG}_{\bm \theta})$ is non-trivial due to the difference in Helmholtz free energies {$A_{\bm \theta}^\mathrm{CG} - A^\mathrm{AT}$} (eq. \ref{eq:RE_definition_canonical}), minimizing $S_\mathrm{rel}(U^\mathrm{CG}_{\bm \theta})$ via first order optimizers only requires the computation of its gradient \cite{Chaimovich2011}
\begin{equation}
\label{eq:rel_entropy_gradient}
    \nabla_{\bm \theta} S_\mathrm{rel}(U^\mathrm{CG}_{\bm \theta}) = \beta \left \langle   \nabla_{\bm \theta} U_{\bm \theta}^\mathrm{CG}(\mathbf{M}(\mathbf{r})) \right \rangle_\mathrm{AT} - \beta \left \langle  \nabla_{\bm \theta} U_{\bm \theta}^\mathrm{CG}(\mathbf{R}) \right \rangle_\mathrm{CG}  \ .
\end{equation}
In practice, the first term in eq. \ref{eq:rel_entropy_gradient} is approximated by an average over the AT reference data set. The second term is computationally more expensive as the distribution corresponding to the CG potential needs to be sampled on-the-fly, typically via a CG MD simulation. The gradient $\nabla_{\bm \theta} U_{\bm \theta}^\mathrm{CG}(\mathbf{R})$ can be computed conveniently via automatic differentiation.

\subsection{Linking Force Matching and Relative Entropy Minimization}
A large body of literature has studied the relationship between FM and RE \cite{Rudzinski2011, Chaimovich2011, Kalligiannaki2015, Kalligiannaki2015, Shell2016}, which we reiterate in parts in the following.
Defining the quantity \cite{Kullback1951}
\begin{equation}
    \Phi_{\bm \theta}(\mathbf{R}) = \ln \left( \frac{p^\mathrm{AT}(\mathbf{R})}{p^\mathrm{CG}_{\bm \theta}(\mathbf{R})} \right)
\end{equation}
allows reformulating the optimization objectives of RE (eq. \ref{eq:CG_S_rel}) and FM (eq. \ref{eq:FM_PMF_form}) \cite{Rudzinski2011} to
\begin{equation}
\begin{split}
        S_\mathrm{rel}  &= \int p^\mathrm{AT}(\mathbf{R}) \Phi_{\bm \theta}(\mathbf{R})  \mathrm{d}\mathbf{R} \\
        \chi^2(U^\mathrm{CG}_{\bm \theta}) &= \frac{(k_\mathrm{B} T)^2}{3n} \int p^\mathrm{AT}(\mathbf{R}) ||\nabla_{\mathbf{R}} \Phi_{\bm \theta}(\mathbf{R})||^2  \mathrm{d}\mathbf{R} + \chi^2(U^\mathrm{PMF})  \ .
\end{split}
\end{equation}
Hence, both FM and RE minimize a functional of $\Phi_{\bm \theta}(\mathbf{R})$. Differences in the learned CG potential result from minimizing an average of $\Phi_{\bm \theta}(\mathbf{R})$ in RE compared to an average of $||\nabla_{\mathbf{R}} \Phi_{\bm \theta}(\mathbf{R})||^2$ in FM.

Thus far, the comparison of FM and RE has generally focused on the case of finite model capacity and infinite AT data: 
FM reaches the minimum of $\chi^2(U^\mathrm{CG}_{\bm \theta})$ (eq. \ref{eq:FM_PMF_form}) if $U^\mathrm{CG}_{\bm \theta}$ is the projection of $U^\mathrm{PMF}$ onto the function space spanned by the CG potential basis set \cite{Noid2008b, Mullinax2009}. However, the resulting $U^\mathrm{CG}_{\bm \theta}$ is not guaranteed to reproduce any AT correlation function mapped to CG coordinates \cite{Rudzinski2011}.
In contrast, the CG potential that minimizes $S_\mathrm{rel}(U^\mathrm{CG}_{\bm \theta})$ is guaranteed to reproduce all mapped AT structural correlation functions that are conjugate to basis functions of the CG potential \cite{Chaimovich2011}. For example, if the CG potential includes a flexible parametrization of pairwise interactions, the radial distribution function of the mapped AT system will be matched.

\subsection{Finite data size effects}
In the following, we compare FM and RE in the context of NN potentials, where we expect a reduced impact of the finite functional basis set, but a larger contribution from finite data size effects.
We assume the common case of a data set generated by an equilibrium AT MD simulation.
Consequently, the data set primarily contains states in potential energy minima, but rarely high energy states \cite{Herr2018}.
This gives rise to two issues in training CG NN potentials via FM: the difficulty of obtaining a globally accurate FES \cite{Chen2021, Kohler2022} and the reliance on prior potentials \cite{Wang2019, Husic2020} (discussed in the next section). 

\begin{figure}
    \centering
    \includegraphics{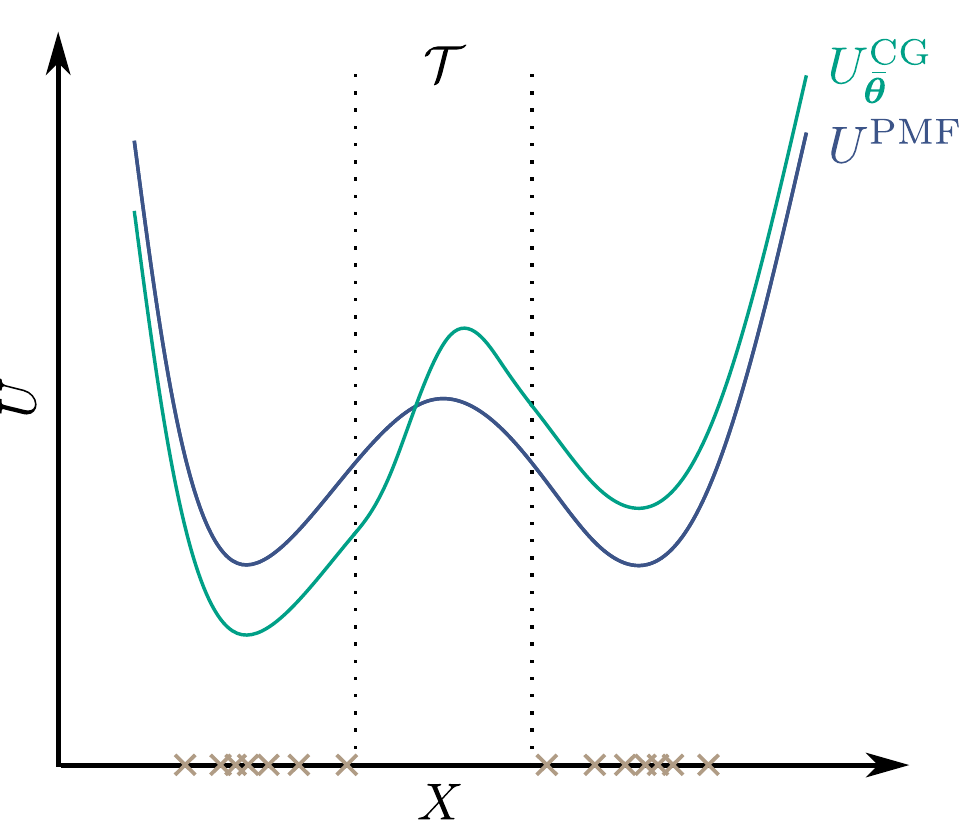}
    \caption{Coarse-graining thought experiment. If the atomistic data set (brown crosses) contains no states within the transition region $\mathcal{T}$, a candidate potential $U^\mathrm{CG}_{\bar{\bm \theta}}$, whose shape only differs from the potential of mean force $U^\mathrm{PMF}$ within $\mathcal{T}$, yields the same force matching validation loss as $U^\mathrm{PMF}$ despite resulting in a different coarse-grained distribution $p^\mathrm{CG}_{\bar{\bm \theta}}(\mathbf{R}) \neq p^\mathrm{AT}(\mathbf{R})$.}
    \label{fig:PMF_Sketch}
\end{figure}

Inaccurate FESs can be caused by sensitivity of the learned potential to errors in sparsely sampled transition regions, as recently hypothesized \cite{Kohler2022}.
We illustrate this idea through a thought experiment, where a system is coarse-grained to a 1D CG coordinate $X$ (fig. \ref{fig:PMF_Sketch}).
We consider a specific CG potential $U^\mathrm{CG}_{\bar{\bm \theta}}$ that differs from $U^\mathrm{PMF}$ within the transition region $\mathcal{T}$. Outside $\mathcal{T}$, $U^\mathrm{CG}_{\bar{\bm \theta}}$ is only shifted with respect to $U^\mathrm{PMF}$. If we assume that the AT data set does not contain any states within $\mathcal{T}$, the validation FM loss of $U^\mathrm{CG}_{\bar{\bm \theta}}$ is identical to the validation FM loss of $U^\mathrm{PMF}$, given that the forces outside $\mathcal{T}$ are identical.
However, the probabilities of samples generated by both potentials differ, e.g. $U^\mathrm{CG}_{\bar{\bm \theta}}$ preferentially samples the left minimum.
FM needs to infer the free energy difference between minima by integrating the mean-force along the transition path, which is unavailable in this thought experiment.
In real-world applications, this mean-force integral is determined by few and noisy \cite{Wang2019} transition states in the AT data set, which explains the reported difficulty in reproducing the correct relative sampling probabilities of different meta-stable states \cite{Chen2021, Kohler2022}.
Since transition states are comparatively rare in the training data, they only have a small impact on the FM validation loss \cite{Herr2018}. Therefore, the FM validation loss is not a useful metric to assess the global quality of the FES \cite{Chen2021}.

In contrast, the incorrect CG distribution $p^\mathrm{CG}_{\bar{\bm \theta}}(\mathbf{R})$ generated by $U^\mathrm{CG}_{\bar{\bm \theta}}$ results in a large $S_\mathrm{rel}$ (eq. \ref{eq:CG_S_rel}). Thus, RE minimization will adjust the potential such that both meta-stable configurations are sampled equally, matching $U^\mathrm{PMF}$ where AT data is available. This is consistent with the interpretation that optimizing $S_\mathrm{rel}(U^\mathrm{CG}_{\bm \theta})$ minimizes the difference between the potential energy surfaces of the AT and CG models \cite{Chaimovich2010}, i.e.
\begin{equation}
\label{eq:S_rel_delta}
    S_\mathrm{rel}(U^\mathrm{CG}_{\bm \theta}) = \ln \left\langle e^{\Delta_{\bm \theta}(\mathbf{r}) - \langle \Delta_{\bm \theta}(\mathbf{r}) \rangle_\mathrm{AT}} \right\rangle_\mathrm{AT} \quad \mathrm{with} \quad \Delta_{\bm \theta}(\mathbf{r}) \equiv \beta[U^\mathrm{AT}(\mathbf{r}) - U^\mathrm{CG}_{\bm \theta}(\mathbf{M}(\mathbf{r}))] \ ,
\end{equation}
where a constant offset between the potential energy surfaces is captured by $\langle \Delta_{\bm \theta}(\mathbf{r}) \rangle_\mathrm{AT}$.
In sum, RE is better suited to reproduce the global FES, especially if the phase-space is resolved inhomogeneously by the training data, making RE minimization more data efficient \cite{Kohler2022}.

\subsection{Prior Potentials}
Classical CG potentials typically use physics-based functional forms \cite{Marrink2007, Ingolfsson2014} that enforce qualitatively correct behavior irrespective of the specific parameter values at hand; for example Lennard-Jones interactions encode the Pauli exclusion principle at short distances and van der Waals forces at longer distances.
To encode physically meaningful behavior in a similar way, the flexible functional form of NN potentials can be combined with a physics-informed prior potential $U^\mathrm{prior}(\mathbf{R})$ \cite{Wang2019, Husic2020, Chen2021, Thaler_2021, Kohler2022}:
\begin{equation}
\label{eq:prior_addition}
    U^{\mathrm{CG}}_{\bm \theta}(\mathbf{R}) = U_{\bm \theta}^{\mathrm{NN}}(\mathbf{R}) + U^\mathrm{prior}(\mathbf{R}) \ .
\end{equation}
In this formulation, training the NN potential $U_{\bm \theta}^{\mathrm{NN}}(\mathbf{R})$ can be interpreted as $\Delta$-learning \cite{Ramakrishnan2015, Shen2018, Boselt2021} with respect to $U^\mathrm{prior}(\mathbf{R})$ \cite{Wang2019}.

Note that the role of $U^\mathrm{prior}(\mathbf{R})$ differs significantly for FM compared to RE minimization:
Since the data set is obtained via physically sound principles in an AT MD simulation, it does not contain any unphysical configurations such as overlapping particles. In such unphysical regions of phase-space, the CG NN potential therefore operates in the extrapolation regime and can easily predict short-range attraction instead of physically sound repulsion.
For FM, a well-chosen $U^\mathrm{prior}(\mathbf{R})$ therefore enforces qualitative correct predictions outside the training data to drive the CG MD simulation back into the AT data distribution, where $U_{\bm \theta}^{\mathrm{NN}}(\mathbf{R})$ is accurate.
Hence, FM requires careful selection of $U^\mathrm{prior}(\mathbf{R})$ given that weak choices can lead to unphysical CG MD simulation results \cite{Wang2009}.

In contrast, a strong deviation of $p^\mathrm{CG}_{\bm \theta}(\mathbf{R})$ from $p^\mathrm{AT}(\mathbf{R})$ caused by an unphysical trajectory leads to a large $S_\mathrm{rel}(U^\mathrm{CG}_{\bm \theta})$, which can be corrected by the optimizer during training.
Rather than stabilizing the application CG MD simulation, the role of $U^\mathrm{prior}(\mathbf{R})$ in RE minimization is to speed-up training convergence. Without a prior, physical principles need to be learned from the AT reference data, which significantly increases the number of update steps until convergence \cite{Thaler_2021}.

\subsection{Finite time step effects}
So far, we have implicitly assumed ideal sampling of the Boltzmann distribution corresponding to a specific potential, i.e., assuming an infinitesimal MD simulation time step $\Delta t$. However, in practice, the AT distribution $p^{AT}_{\Delta t_\mathrm{AT}}(\mathbf{r})$ results from the time step-dependent shadow Hamiltonian \cite{Toxvaerd1994} of the reference AT simulation \cite{Kohler2022} (with the shadow temperature \cite{Toxvaerd2013} representing the conserved quantity in the NVT ensemble). Thus, RE learns a CG potential whose shadow Hamiltonian yields a CG distribution $p^\mathrm{CG}_{\bm \theta, \Delta t_\mathrm{CG}}(\mathbf{R})$ that approximates $p^{AT}_{\Delta t_\mathrm{AT}}(\mathbf{R})$. Consequently, assuming infinite data and model capacity, the optimal RE potential differs from the true PMF as a function of $\Delta t_\mathrm{AT}$ and $\Delta t_\mathrm{CG}$.
On the other hand, FM models train on a data set of forces computed from the true AT potential. Hence, the ideal FM potential equals the true PMF, but the resulting CG distribution $p^\mathrm{CG}_{\bm \theta, \Delta t_\mathrm{CG}}(\mathbf{R})$ will differ from the analytic $p^{AT}(\mathbf{R})$ as a function of the production time step $\Delta t_\mathrm{CG}$.

\subsection{Neural Network Potential}
We use our previously published implementation \cite{Thaler_2021} of the graph NN DimeNet++ \cite{Klicpera2020, Klicpera2020b} as $U_{\bm \theta}^\mathrm{NN}$ with graph cut-off radius $r_\mathrm{cut} = 0.5$ nm. We set all hyperparameters to the default values of the original implementation \cite{Klicpera2020b}, except for embedding sizes, which we reduce by factor 4.
With the default of 4 interaction blocks, DimeNet++ captures up to 8-body correlations \cite{Batatia2022b} as angles are a direct input quantity that already capture 3-body properties. This high-body interaction capacity promises highly accurate approximations to the PMF.

\section{Results}
\subsection{Liquid Water}
We choose the classical benchmark problem of CG liquid water to test FM and RE in a setting where AT reference data is abundantly available. We generate a 10~ns AT trajectory of the TIP4P/2005 \cite{Abascal2005} water model at a temperature $T_\mathrm{ref}=298$~K, which we subsample to retain a state every 1~ps. Each state consists of a cubical simulation box of length $l=3.129$~nm containing 1000 water molecules. 
The first 8~ns are used for training, the subsequent 0.8~ns for validation and the last 1.2~ns are retained as a test set.

We select a CG mapping, where each water molecule is mapped to a CG particle located at its center of mass.
To the DimeNet++ $U_{\bm \theta}^{\mathrm{NN}}(\mathbf{R})$ we add the pairwise repulsive part of the Lennard-Jones potential as prior
\begin{equation}
\label{eq:prior_water}
    U^\mathrm{prior}(\mathbf{R}) = \sum_{i=1}^{N_\mathrm{pair}} \epsilon \left( \frac{\sigma}{d_i} \right)^{12} \ ,
\end{equation}
where we sum over all $N_\mathrm{pair}$ pairs with distance $d_i < r_\mathrm{cut}$ (eq. \ref{eq:prior_addition}). Analogous to our previous work \cite{Thaler_2021}, we choose $\epsilon = 1\ \mathrm{kJ}/\mathrm{mol}$ and $\sigma = 0.3165$~nm, which is the length scale of the SPC \cite{Berendsen1981} water model.

The FM model is trained for 100 epochs with a batch size of 10 states (for loss curves, see supplementary fig. 1). We select the model with the smallest validation loss, which is computed after each training epoch. The validation set is exclusively used in FM for this purpose, giving FM a small advantage over RE in terms of data usage.
We train the RE model for 300 update steps. To sample $p^\mathrm{CG}_{\bm \theta}(\mathbf{R})$ during training, we run 70~ps simulations, including 5~ps of equilibration, with a time step of 2~fs. Each trajectory starts from the last state of the previous trajectory to reduce equilibration time. For more technical details, see supplementary methods 1.

\begin{figure}
    \centering
    \resizebox{\linewidth}{!}{\includegraphics{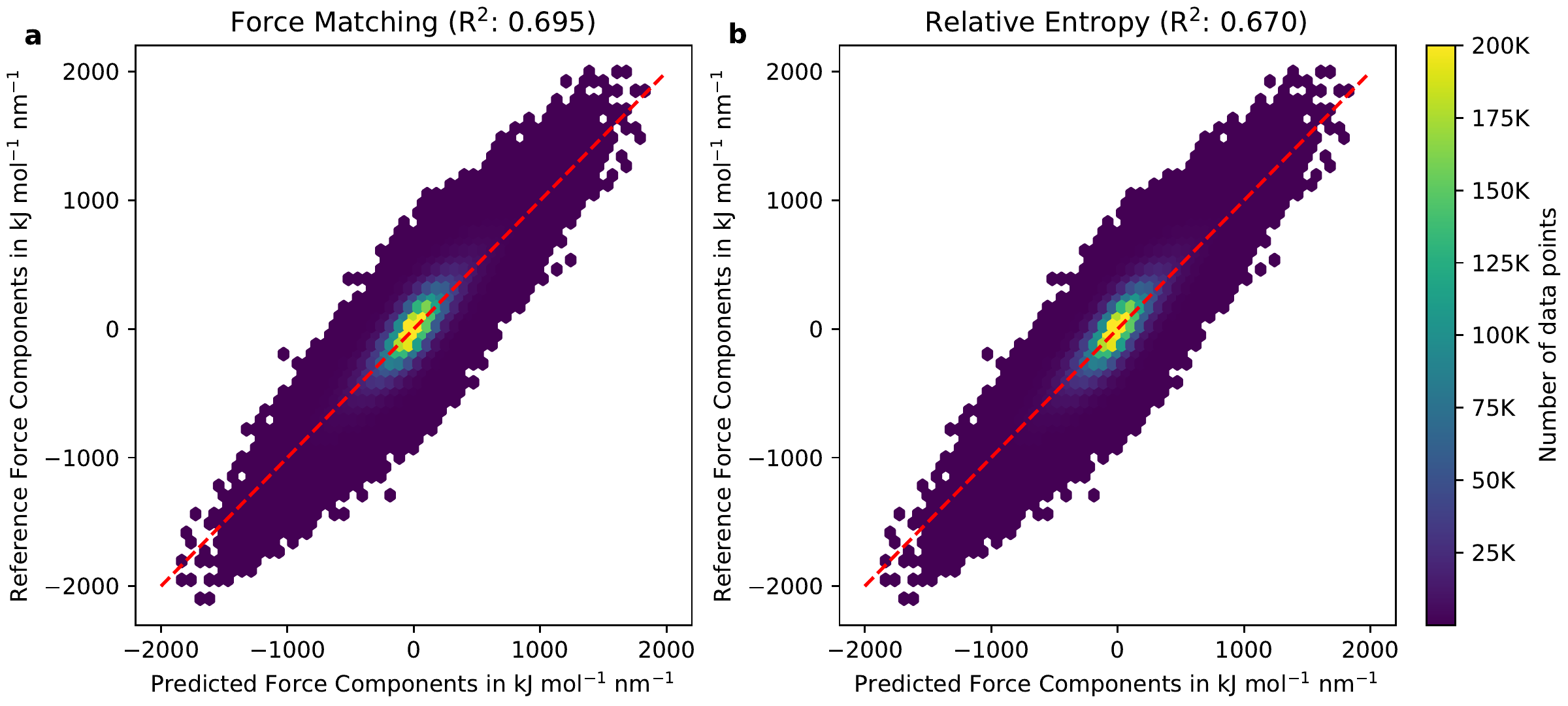}}
    \caption{Liquid water force predictions on test data. Each data point corresponds to a predicted force component for a coarse-grained particle in the test data set compared to its atomistic reference for models trained via ($\mathbf{a}$) force matching  and ($\mathbf{b}$) relative entropy minimization.}
    \label{fig:water_scatter}
\end{figure}

We evaluate the quality of force predictions of both trained models based on the held-out test data set (fig. \ref{fig:water_scatter}).
Compared to AT NN potentials, predicted forces exhibit larger errors due to the noise resulting from the non-injective CG mapping (eq. \ref{eq:FM_PMF_form}).
The FM model yields slightly better force predictions ($R^2 = 0.695$) than the RE model ($R^2 = 0.670$). Apart from possible overfitting of the FM model onto forces, this result likely stems from finite time step effects discussed above \cite{Kohler2022}: The reference forces are computed from the true AT potential, which is consistent with the optimization objective of the FM method. In contrast, the RE potential needs to account for the shadow Hamiltonians of the AT and CG simulations.

\begin{figure}
    \centering
    \resizebox{\linewidth}{!}{\includegraphics{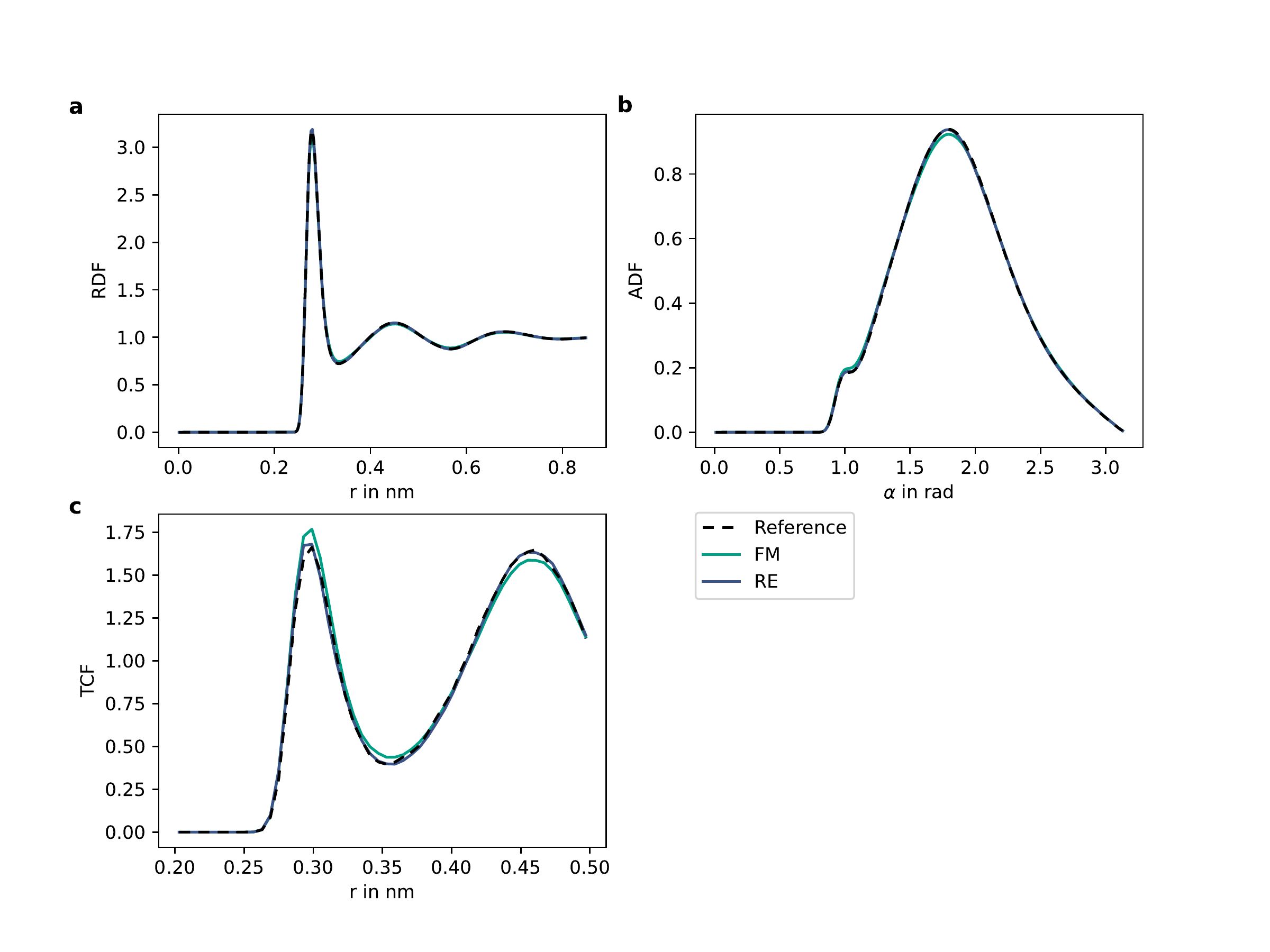}}
    \caption{Structural correlation functions in liquid water. Resulting ($\mathbf{a}$) radial (RDF) and ($\mathbf{b}$) angular distribution function (ADF) \cite{Soper2008} as well as ($\mathbf{c}$) equilateral triplet correlation function (TCF) \cite{Bildstein1994, Dhabal2014} of models trained via force matching (FM) and relative entropy (RE) minimization compared to the atomistic reference.
    \label{fig:water-structure}
    }
\end{figure}

To test the capabilities of the models in an application context, we perform CG MD simulations with a time step of $\Delta t_\mathrm{CG} = 2$ fs and a trajectory length of 1.1 ns, where the first 0.1 ns are discarded for equilibration. We compute the radial (RDF) and angular distribution function (ADF) \cite{Soper2008} as well as the equilateral triplet correlation function (TCF) \cite{Baranyai1990, Bildstein1994, Dhabal2014} to assess different structural correlations in the generated CG distribution. The RE model matches the AT references to the line thickness (fig. \ref{fig:water-structure}). 
This is in line with theoretical expectations that RE reproduces all structural correlation functions for which conjugate terms in the CG potential exists \cite{Chaimovich2011}. FM is in better agreement with the AT reference pressure $p_\mathrm{ref} = -6.2 \ \mathrm{MPa}$ ($p_\mathrm{FM} = 212.2 \ \mathrm{MPa}$, $p_\mathrm{RE} = 311.0 \ \mathrm{MPa}$) at the expense of slightly larger errors in structural correlation functions.
These results are insensitive to the specific choice of prior potential, which we tested by selecting a softer prior $\left(\frac{\sigma}{d}\right)^6$ (eq. \ref{eq:prior_water}; supplementary fig. 2).

Additionally, we compare the DimeNet++ model to a classical 2-body cubic spline model. The spline model is computationally inexpensive compared to the DimeNet++ model (184.5 ps/min versus 7.3 ps/min), at the expense of reduced accuracy: In contrast to the FM spline model, the RE spline model matches the target RDF (supplementary fig. 3) - reproducing literature results \cite{Chaimovich2009, Ruhle2009}. However, as expected, both models fail to match 3-body correlations. Adding 3-body terms improves those classical models \cite{Scherer2018}, but the accuracy still remains limited compared to the DimeNet++ model.
Overall, these results suggest that the difference between models obtained via FM and RE tends to increase for decreasing adequacy of the functional basis set for a given system - in line with previous computational studies \cite{Kalligiannaki2016}.

\begin{figure}
\includegraphics[width=\textwidth]{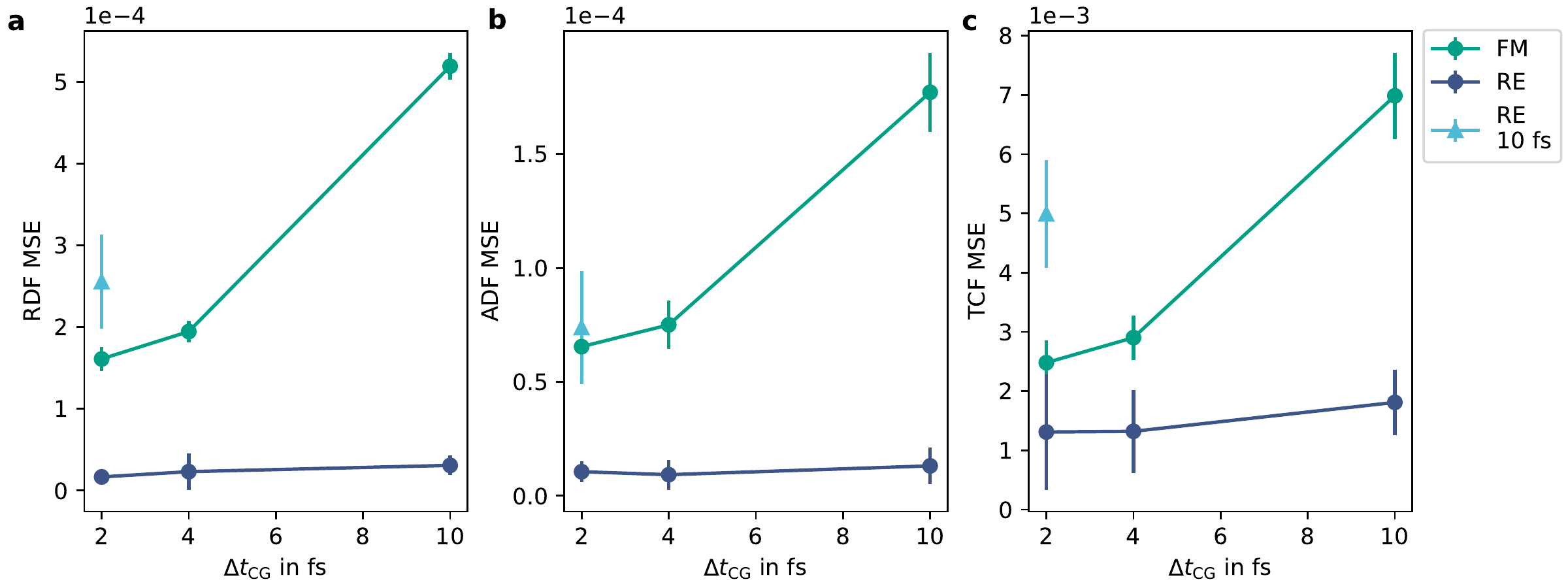}
  \caption{Time step variation. Mean squared error (MSE) of resulting ($\mathbf{a}$) radial (RDF) and ($\mathbf{b}$) angular distribution function (ADF) as well as ($\mathbf{c}$) equilateral distribution function (TCF) for different time step sizes $\Delta t_\mathrm{CG}$ in subsequent molecular dynamics simulations. The plotted mean and standard deviation values are computed from 5 models with different random seeds for neural network parameter initialization and velocity distribution of the initial simulation state. For force matching (FM), the same 5 models are used for different simulation time steps. For relative entropy (RE) minimization, the models are re-trained such that the training time step matches $\Delta t_\mathrm{CG}$. An exception are models trained with 10~fs, which are additionally run with $\Delta t_\mathrm{CG} = 2$~fs (light blue).}
  \label{fig:time_step_variation}
\end{figure}

Given that computational speed-up is the principal motivation for CG modeling, we evaluate trained FM and RE models for the real-world case of larger production CG MD time step sizes $\Delta t_\mathrm{CG}$ (fig. \ref{fig:time_step_variation}). 
The resulting MSE values of FM models increase significantly for larger $\Delta t_\mathrm{CG}$, which we attribute to increased time integration errors. Presumably, the impact of the CG shadow Hamiltonian becomes noticeable for FM CG NN potentials as errors from an incomplete basis set and finite data size effects are small in this problem.
By contrast, the MSE values of RE models increase only slightly for larger $\Delta t_\mathrm{CG}$ when using the same time step during training.
Given that RE optimizes the empirical CG distribution $p^\mathrm{CG}_{\bm \theta, \Delta t_\mathrm{CG}}(\mathbf{R})$, we presume that the RE potential learns to correct for the time step-dependent terms in the CG shadow Hamiltonian.
To test this hypothesis, we apply the RE models trained with 10~fs in CG MD simulations with $\Delta t_\mathrm{CG} = 2$~fs. In line with our hypothesis, this combination yields larger MSE values than RE models with consistent time steps (fig. \ref{fig:time_step_variation}): If the RE model learns to correct large time step-dependent terms in the shadow Hamiltonian, this biases CG simulations that exhibit only small time integration errors.
Hence, RE minimization provides a means to mitigate the accuracy degradation of larger production time steps $\Delta t_\mathrm{CG}$.

\subsection{Alanine Dipeptide}
Alanine Dipeptide \cite{Pettitt1985, Tobias1992} is a standard problem to benchmark CG methods in reconstructing a FES with multiple meta-stable states. We generate a 100~ns AT reference trajectory at $T_\mathrm{ref} = 300$~K from which a state is retained every 0.2~ps, resulting in $5 \cdot 10^5$ data points. The training data set consists of the first 80~ns, the FM validation set of the subsequent 8~ns and the final 12~ns form the test set.
We select a CG mapping that retains all 10 heavy atoms of alanine dipeptide, but drops hydrogen atoms and water molecules.
The CG particles representing $\mathrm{CH}_3$, CH and C are encoded as different particle types.
Following the $\Delta$-learning ansatz in eq. \ref{eq:prior_addition}, we select a prior potential
\begin{equation}
    \label{eq:ala_prior}
    \begin{split}
    & U^\mathrm{prior}(\mathbf{R}) = \sum_{i=1}^{N_\mathrm{bonds}} U^\mathrm{harmonic}(b_i) + \sum_{j=1}^{N_\mathrm{angles}} U^\mathrm{harmonic}(\alpha_j) + \sum_{k=1}^{N_\mathrm{dihedrals}} U^\mathrm{proper}(\omega_k) \\
    & U^\mathrm{harmonic}(x_i) = \frac{k_\mathrm{B}T}{2\mathrm{Var}[{x_i}]}(x_i - \langle x_i \rangle_\mathrm{AT})^2 \quad \mathrm{with} \quad  \mathrm{Var}[{x_i}] = \langle (x_i - \langle x_i \rangle_\mathrm{AT})^2 \rangle_\mathrm{AT}  \\
    & U^\mathrm{proper}(\omega_i) = k_{\omega}(1+\cos{n\omega_i-\omega_0}) \ ,
    \end{split}
\end{equation}
where we sum over all $N_\mathrm{bonds}$ harmonic bonds with bond lengths $b_i$, all $N_\mathrm{angles}$ harmonic angles with triplet angles $\alpha_j$ and all $N_\mathrm{dihedrals}$ proper dihedral angles $\omega_k$.
The dihedral force constant $k_{\omega}$, the multiplicity $n$, and the phase constant $\omega_0$ are taken from the AMBER03 \cite{Duan2003} force field.

We train the FM model for 100 epochs with a batch size of 500 states and select the model that yields the smallest validation loss (for loss curves, see supplementary fig. 4). For RE training, we sample the CG distribution through 50 vectorized CG MD simulations starting from different initial states, which improves computational efficiency on GPUs. Each parallel simulation generates a 1~ns trajectory, of which 5~ps are discarded for equilibration. The simulations restart from the last obtained state of the previous trajectory and we train the RE model for 300 updates.
Additional technical details are available in supplementary methods 2.

First, we compare the force prediction quality of the FM and RE models on the test data set (scatter plots in supplementary fig. 5). Analogous to the liquid water example, the FM model yields better force predictions ($R^2 = 0.780$) than the RE model ($R^2 = 0.717$). However, in the case of Alanine Dipeptide, we are mostly interested in an accurate reproduction of the FES. Hence, we sample 100 ns CG trajectories with the trained RE and FM models such that the number of generated states equals the AT reference data.

The resulting 2D density histograms of the dihedral angles $\phi$ and $\psi$ \cite{Pettitt1985, Tobias1992} are shown in fig. \ref{fig:ala_density} (corresponding FESs in supplementary fig. 6).
\begin{figure}
    \centering
  \includegraphics[width=\textwidth]{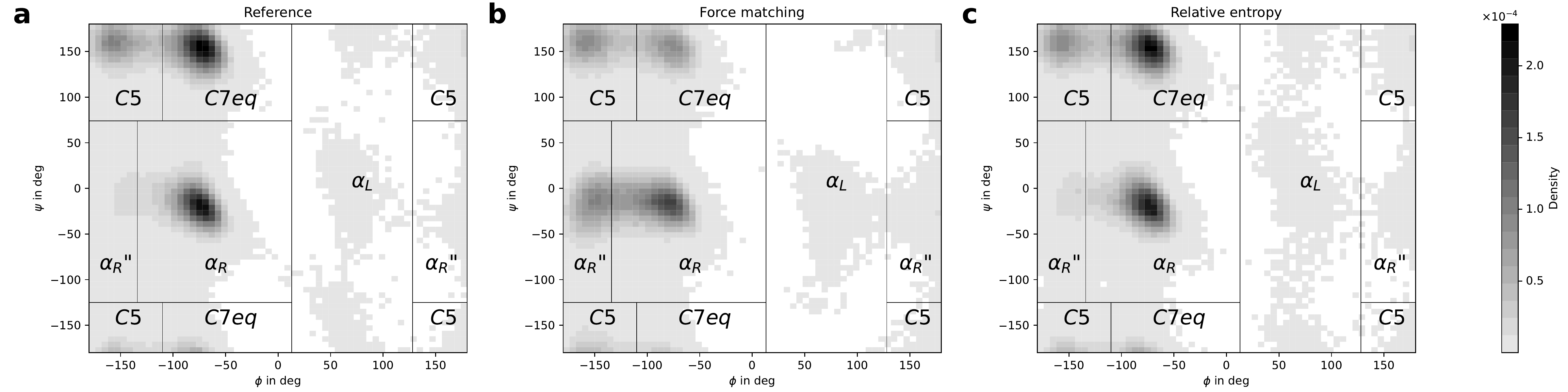}
  \caption{Ramachandran diagrams. Resulting density histograms of the dihedral angles $\phi$ and $\psi$ from ($\mathbf{a}$) the AT reference simulation and from the CG models trained via ($\mathbf{b}$) force matching  and ($\mathbf{c}$) relative entropy minimization.}
  \label{fig:ala_density}
\end{figure}
The FESs obtained via the Dimenet++ potential compare favourably to previously reported results with a classical generalized Born \cite{Chen2008} implicit solvent model of Alanine Dipeptide \cite{Chen2021}.
The Ramachandran diagram of the RE model matches the AT reference well, but the FM model oversamples the $\alpha_R^\mathrm{''}$ configuration.
The 1D projections of the dihedral density are shown in fig. \ref{fig:ala_1d_density}.
\begin{figure}
    \centering
    \includegraphics[width=\textwidth]{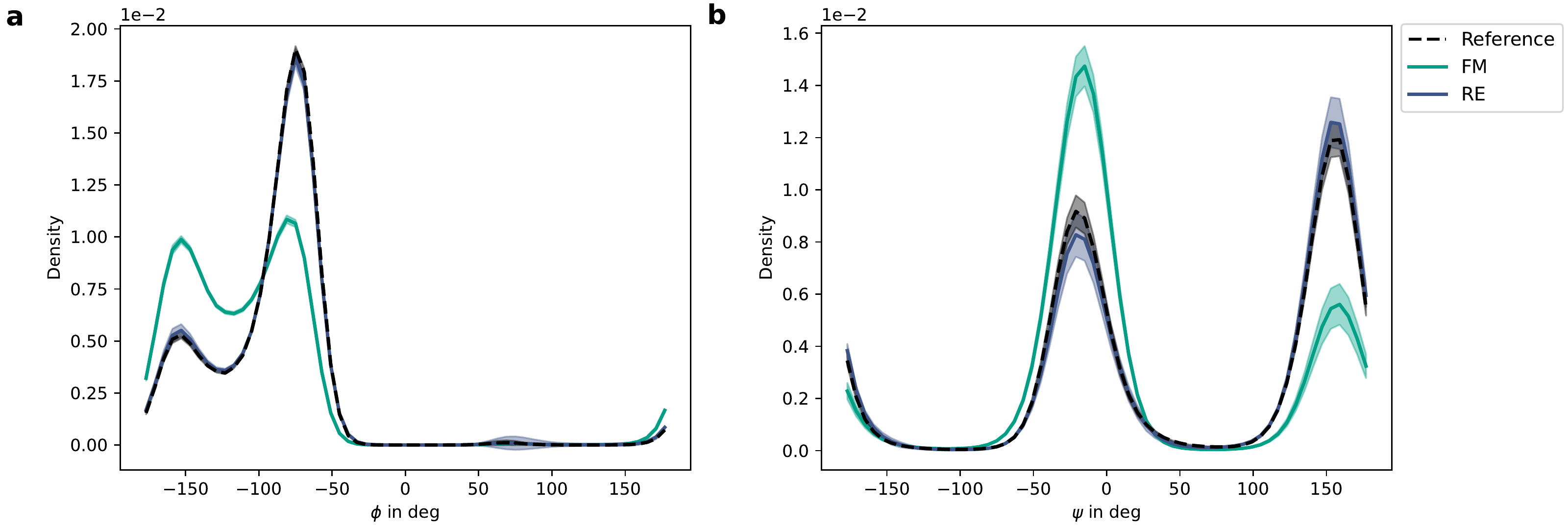}
    \caption{Dihedral angle density. Distribution of dihedral angles ($\mathbf{a}$) $\phi$ and ($\mathbf{b}$) $\psi$ as predicted from the CG models trained via force matching (FM) and relative entropy (RE) minimization, compared to the atomistic reference. The mean and standard deviation (shaded area) are computed from 50 trajectories of 100 ns length.} 
    \label{fig:ala_1d_density}
\end{figure}
Despite the smaller test set error with respect to forces, the FM model fails to accurately reproduce the ratio of meta-stable states.
This result supports the notion that the FM validation error is not a useful metric to judge the global quality of the learned FES \cite{Kohler2022}.

\begin{figure}
    \centering
\includegraphics[width=.6\textwidth]{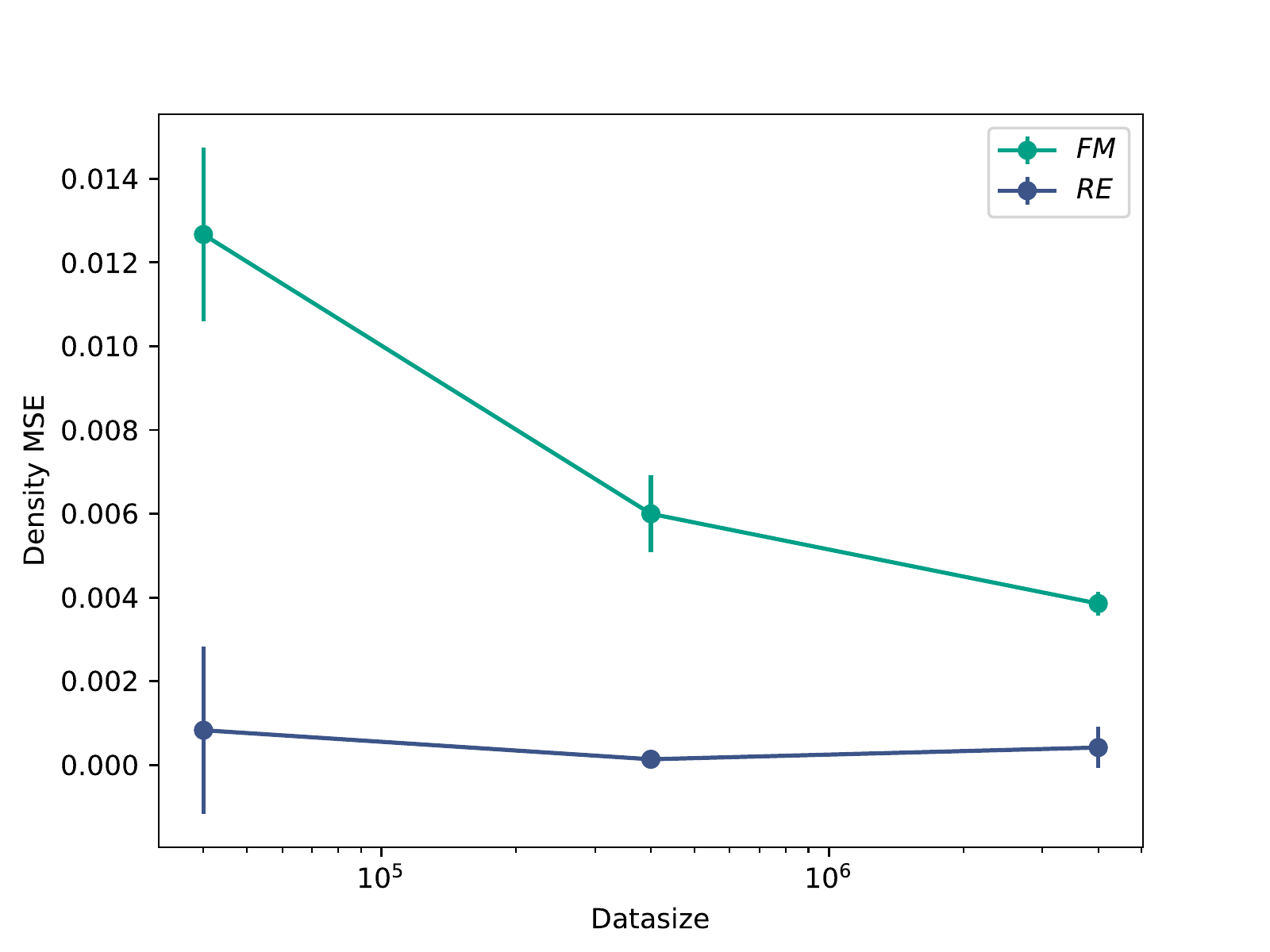}
  \caption{Training data variation. Mean squared error (MSE) of the $\phi$ - $\psi$ dihedral density histograms of force matching (FM) and relative entropy (RE) minimization models for varying training data sizes. The mean and standard deviation values are computed from 50 trajectories of 100 ns length.}
  \label{fig:ala_data_variation}
\end{figure}

Assuming that insufficient resolution of transition regions caused the suboptimal FES of the FM model, increasing the amount of training data should improve the FES: We generate a 1 $\mu$s AT trajectory, increasing the amount of training data by factor 10.
Using this data set, the error of the FM model decreases as expected, but is still inferior to the RE model (fig. \ref{fig:ala_data_variation}). Note that numerical errors of the CG simulation also contribute to the FM MSE, which cannot be reduced by enlarging the training data set.
To test the data requirement limits of RE, we reduce the 100~ns training data set by factor 10. In this case, RE still results in a smaller error than FM with the largest data set despite using only 1\% of the training data, which highlights the data efficiency of RE for reproducing the FES.

By combining FM and RE, we aim to exploit their respective strengths - data efficiency of RE and computational inexpensiveness of FM. Based on the 100~ns data set, we optimize the FM potential by additional RE updates. 
\begin{figure}
    \centering
    \includegraphics[width=\textwidth]{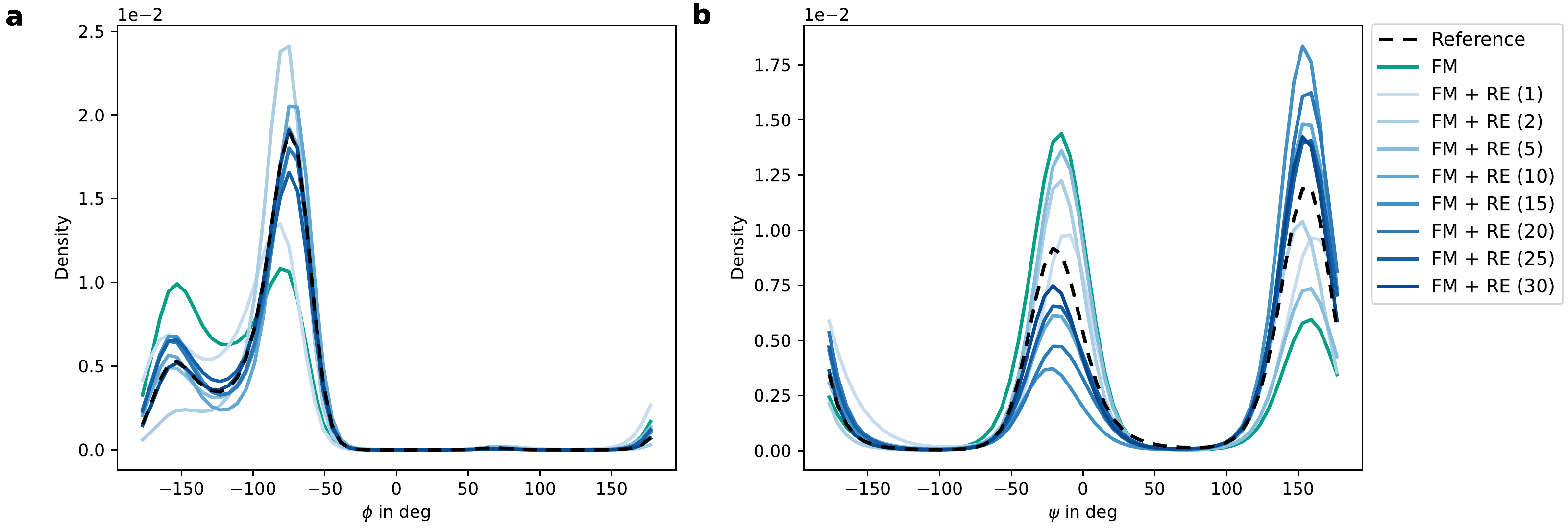}
    \caption{Convergence of relative entropy correction steps. ($\mathbf{a}$) $\phi$ and ($\mathbf{b}$) $\psi$ dihedral angle distributions corresponding to potentials obtained by different numbers of relative entropy (RE) updates when being initialized to the force matching (FM) potential. The lines represent the mean computed from 50 trajectories of 100 ns length.} 
    \label{fig:ala_pretrain}
\end{figure}
As depicted in fig. \ref{fig:ala_pretrain}, few RE updates are sufficient to significantly improve the FES. With 30 updates, the obtained dihedral densities are comparable to the randomly initialized 300 update RE model (fig. \ref{fig:ala_1d_density}) and significantly better than a randomly initialized 30 update RE model (supplementary fig. 7). Consequently, initializing RE minimization with the FM model allows to reduce the number of necessary RE updates significantly.

\begin{figure}
    \centering
    \includegraphics[width=\textwidth]{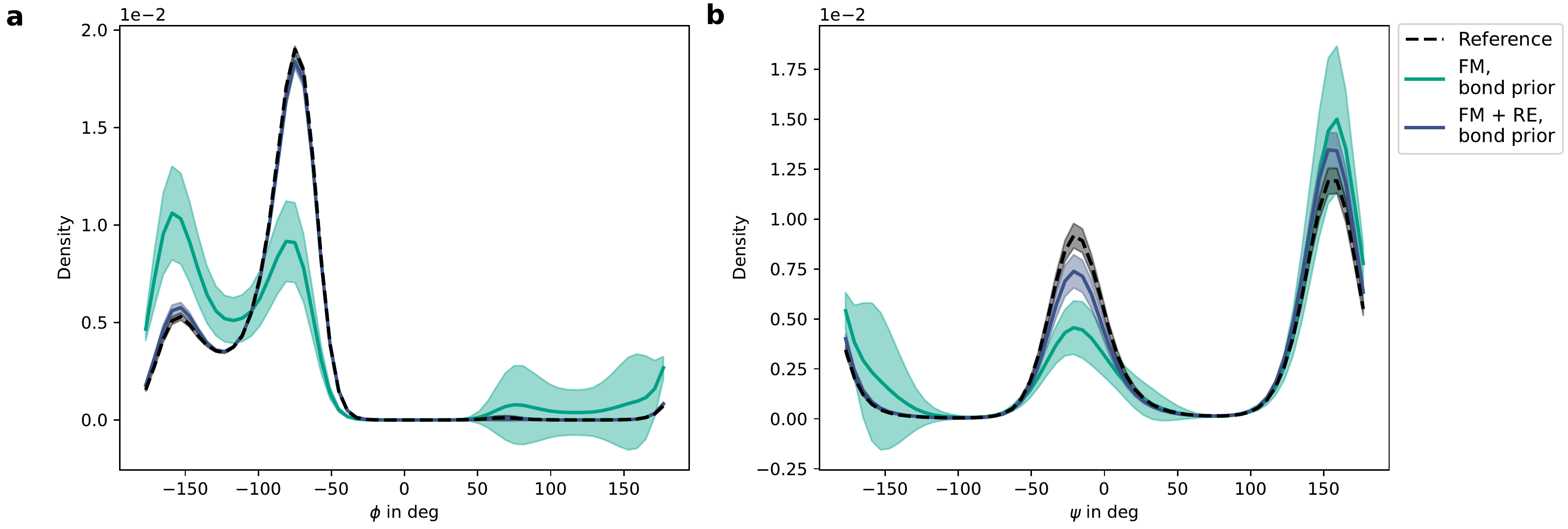}
    \caption{Prior potential ablation. Distribution of dihedral angles ($\mathbf{a}$) $\phi$ and ($\mathbf{b}$) $\psi$ resulting from the force matching (FM) model when only considering bonds in the prior potential. These are compared to the atomistic reference and to a model that optimizes the FM potential via 300 relative entropy (RE) update steps. The mean and standard deviation (shaded area) are computed from 50 trajectories of 100 ns length.} 
    \label{fig:ala_bond_prior}
\end{figure}

Finally, we test the robustness of both methods with respect to prior potentials by considering only harmonic bonds in $U^\mathrm{prior}(\mathbf{R})$ (eq. \ref{eq:ala_prior}). In this case, the FM potential significantly oversamples $\alpha_L$ configurations (fig. \ref{fig:ala_bond_prior}, supplementary fig. 8), despite superior validation force predictions ($R^2 = 0.762$) compared to the reference RE model above.
Conversely, when optimizing the resulting FM model via 300 additional RE update steps, the dihedral density is in close agreement with the AT reference. Hence, RE minimization also helps correcting weak choices of prior potentials.

\section{Discussion and Conclusion}
In this work, we have demonstrated the effectiveness of training CG NN potentials via the RE minimization scheme: For water, the difference between FM and RE minimization is significantly reduced when training CG NN potentials compared to classical 2-body CG potentials \cite{Scherer2018}. This is expected as the learned potential converges towards the PMF for increasing model capacity with both methods, given a sufficient amount of data \cite{Noid2008, Chaimovich2011, Rudzinski2011}. 
For alanine dipeptide, RE results in a more accurate FES than FM.
The discrepancy in the FES increases for decreasing quality of the prior to the point that the FM CG NN potential is no longer competitive with classical CG models.
Sampling the CG model during training probes its robustness with respect to data generated on-the-fly. As a consequence, the RE training scheme can recognize and correct undesired model properties, which reduces the sensitivity on the prior potential.

For liquid water, RE allows larger time steps in subsequent CG MD simulations without compromising accuracy. We presume that RE is able to learn to correct the time integration error of the underlying CG MD simulation by training directly on the sampled CG distribution. 
Given that computational speed-up is the primary objective of CG modeling, a larger simulation time step is as important as an accurate approximation of the PMF. Consequently, CG NN potentials trained via RE may reach larger time scales in production CG simulations with less impact from time integration errors.

The advantages of RE minimization come at the cost of increased computational effort during training, which can however be reduced:
As demonstrated in this work, pre-training via FM is a computationally efficient way to reduce the number of necessary RE updates through a better parameter initialization.
Furthermore, histogram reweighting \cite{Zwanzig1954, Chipot2007, Norgaard2008, Li2011, Carmichael2012, Shell2016}, frequently used in RE minimization \cite{Shell2008, Chaimovich2011, Bottaro2013}, is also applicable to NN potentials \cite{Thaler_2021}. Reweighting allows previously generated trajectories to be reused, increasing the number of gradient descent steps per trajectory computation.
Additionally, it seems reasonable to increase the trajectory length during the course of the optimization. In the beginning of training, where the model error significantly exceeds the statistical error of trajectories, short trajectories can save compute. Towards the end of training, longer trajectories with reduced statistical noise allow fine-tuning of the model. This scheme matches well with reweighting: Initial trajectories cannot be used for reweighting, irrespective of their length, due to large changes in the potential, while expensive trajectories towards the end of training may be reused for multiple updates.
Moreover, we argue that in the realistic scenario of an expensive AT model with a, by design, orders of magnitude cheaper CG model, the computational bottleneck is AT training data generation rather than CG model optimization.
Finally, NN potential architectures optimized for computational efficiency, such as the Ultra-Fast Force Fields \cite{Xie2021}, are well-suited for CG applications. These architectures allow to capture many-body features of the PMF while reducing the computational overhead of the more expensive DimeNet++ \cite{Klicpera2020b} model used in this work.
Evaluating the computational cost-accuracy trade-off between different computationally efficient CG NN potentials and classical CG models is an interesting avenue of future research.
Additionally, the merits of CG NN potentials should be examined for more complex systems than considered in this work.

The presented results can also be interpreted in terms of data efficiency. In CG applications, an accurate representation of the FES is usually of higher interest than accurate force predictions in energy minima. RE is well suited to reproduce the FES in practice by directly minimizing the difference between the potential energy surfaces of the AT and CG models (eq. \ref{eq:S_rel_delta}).
By contrast, FM requires a sufficient resolution of transition areas to learn a globally accurate FES \cite{Kohler2022}. This requires a large amount of AT training data, which is expensive to obtain.
Long AT MD trajectories do not seem efficient in this regard due to repetitive sampling of energy minima and sparse sampling of high energy states \cite{Herr2018}. Accordingly, enhanced sampling schemes such as Metadynamics \cite{Barducci2011, Bonati2018, Herr2018} or Normal Mode Sampling \cite{Schneider1989, Rupp2015} may improve data efficiency of FM by spreading sampling more evenly across the phase space.

In line with literature on simulation-based optimization schemes for classical CG models \cite{Reith2003, Noid2013}, our results suggest that including MD simulations in the training process can be considered as a means to improve the reliability and accuracy of NN potentials, allowing to address recent concerns about their stability \cite{Stocker2022, Fu2022}.
Active learning NN potentials \cite{Smith2018, Zhang2019}, recognized as a major building block in achieving stable and transferable models \cite{Loeffler2020, Jinnouchi2020, Smith2021}, can be similarly interpreted as an incorporation of MD simulations into the FM training scheme:
Performing MD simulations and screening visited molecular states for high-uncertainty configurations allows to augment the data set iteratively in phase space regions that are reachable by the NN potential but still sparsely represented in the data set.
Alternatively, MD simulations can also be inserted directly into the training pipeline \cite{Ingraham2019, Schoenholz2020, Goodrich2021, Doerr2021, Thaler_2021} using auto-differentiable MD codes \cite{Schoenholz2020, Doerr2021}.
We expect that the benefits of using ML in simulations and, inversely, simulations for ML training will continue to drive the ongoing synthesis of ML and physical simulations in molecular modeling and beyond.

\begin{acknowledgments}
The authors thank Dominik Blechschmidt for contributions to initial feasibility studies.
\end{acknowledgments}

\section*{Data Availability Statement}
The data that support the findings of this study are available from the corresponding author upon reasonable request.
The code to train presented DimeNet++ models via FM and RE is open-sourced at \href{https://github.com/tummfm/relative-entropy}{https://github.com/tummfm/relative-entropy}.

\bibliography{library}

\end{document}